\documentclass[a4paper,10pt,aps,pre,twocolumn,amsmath,amssymb]{revtex4}
\usepackage{graphicx,color}

\newcommand{\ahum}[1]{``#1''}
\newcommand{\eq}[1]{Eq.~(\ref{#1})}
\newcommand{\fig}[1]{Fig.~\ref{#1}}
\newcommand{\sect}[1]{Section~\ref{#1}}
\newcommand{\olcite}[1]{Ref.~\cite{#1}}
\newcommand{\avg}[1]{\langle #1 \rangle}

\newcommand{\tc}{T_{\rm c}}

\begin{document}

\title{A finite-temperature Monte Carlo algorithm for network forming 
materials}

\author{Richard L. C. Vink}
\affiliation{Institute of Theoretical Physics, Georg-August-Universit\"at 
G\"ottingen, Friedrich-Hund-Platz~1, D-37077 G\"ottingen, Germany}

\date{\today}

\begin{abstract} Computer simulations of structure formation in network forming 
materials (such as amorphous semiconductors, glasses, or fluids containing 
hydrogen bonds) are challenging. The problem is that large structural changes in 
the network topology are rare events, making it very difficult to equilibrate 
these systems. To overcome this problem, Wooten, Winer and Weaire 
[Phys.~Rev.~Lett.~{\bf 54}, 1392 (1985)] proposed a Monte Carlo bond-switch 
move, constructed to alter the network topology at every step. The resulting 
algorithm is well suited to study networks at zero temperature. However, since 
thermal fluctuations are ignored, it cannot be used to probe the phase behavior 
at finite temperature. In this paper, a modification of the original bond-switch 
move is proposed, in which detailed balance and ergodicity are both obeyed, 
thereby facilitating a correct sampling of the Boltzmann distribution for these 
systems at any finite temperature. The merits of the modified algorithm are 
demonstrated in a detailed investigation of the melting transition in a 
two-dimensional 3-fold coordinated network. \end{abstract}

\maketitle

\section{Introduction}

Network forming materials are ubiquitous in nature, common examples being 
semiconductors such as silicon and silica, as well as fluids that can form 
hydrogen bonds. What these materials have in common is that their topology on 
short length scales is governed by certain rules. For example, in amorphous 
silicon, most atoms are 4-fold coordinated, the preferred Si-Si bond length 
being $\approx 2.35$~\AA, and the preferred Si-Si-Si bond angle being the 
tetrahedral angle. This complicates molecular dynamics simulations of these 
materials, where the particles spend most of their time thermally fluctuating 
about their equilibrium positions, while large structural changes in the network 
topology are rare. This is particularly cumbersome if well-relaxed amorphous 
networks are needed, i.e.~networks that locally fulfill the bond requirements 
(and thus have a low energy) but where long-ranged order is absent.

\begin{figure}
\begin{center}
\includegraphics[width=0.9\columnwidth]{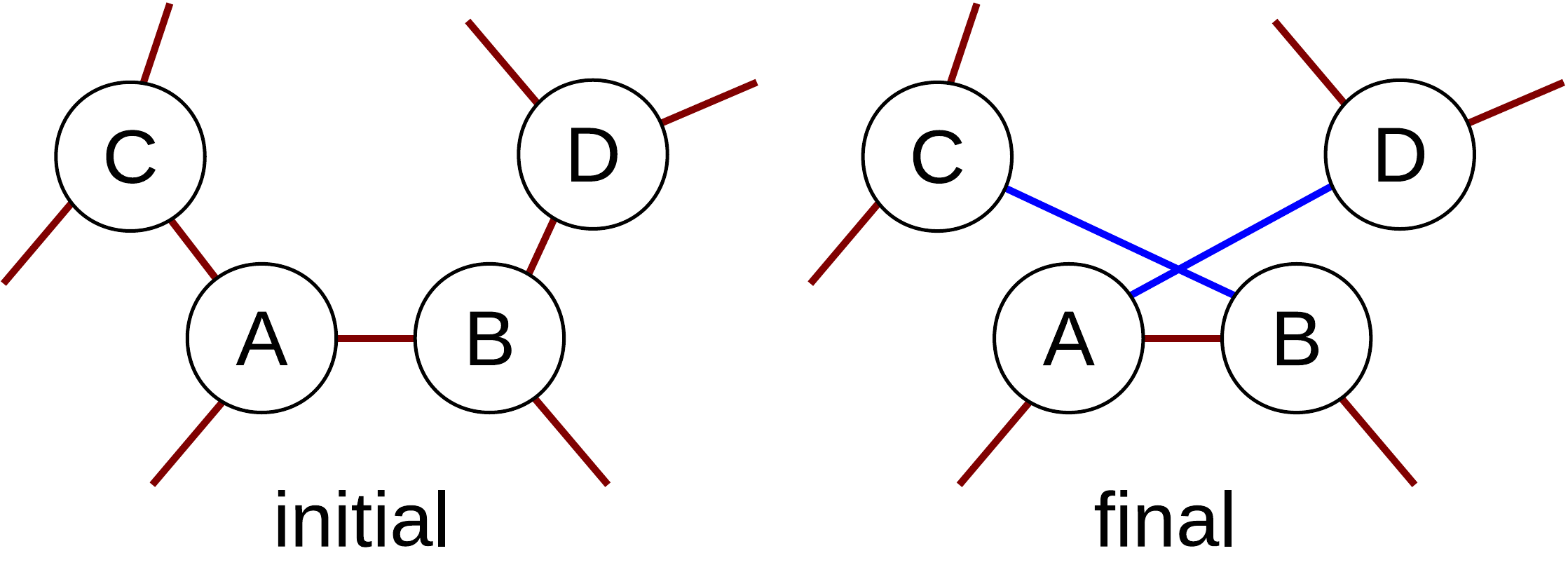}
\caption{\label{fig1} The bond-switch Monte Carlo move of the WWW algorithm.}
\end{center}
\end{figure}

To overcome this problem, Wooten, Winer, and Weaire (WWW) proposed a Monte Carlo 
algorithm using a bond-switch move~\cite{citeulike:3269654}. This move can be 
applied to any system whose potential energy $E$ is defined via a connectivity 
table, i.e.~an explicit list specifying which particles are connected to each 
other by bonds (the prototype example of such a potential is the Keating 
potential~\cite{citeulike:7333784}, see also \eq{eq:keating}). The bond-switch 
move proceeds as shown in \fig{fig1}. First, a cluster $\{A,B,C,D\}$ of four 
particles is selected randomly, containing the bonds $\{AB, AC, BD\}$ with the 
constraint that $A$ may not be bonded to $D$, nor $B$ to $C$. Next, a change in 
the network topology is proposed, whereby the bonds $AC$ and $BD$ are removed 
from the connectivity table, and replaced by two new bonds, $AD$ and $BC$. The 
change is accepted with the Metropolis probability
\begin{equation}\label{eq:www}
 P_{\rm acc} = \min[1, e^{-(E_F' - E_I') / k_BT}],
\end{equation}
with $E_I' \, (E_F')$ the energy before (after) the bond-switch, $T$ the 
temperature, and $k_B$ the Boltzmann constant. The prime ($'$) indicates, and 
this point is crucial, that the energy is to be measured with all the particles 
in the network placed at their equilibrium positions. That is, for a given 
network topology (connectivity table) the particle positions are fixed 
deterministically by energy minimization.

The WWW algorithm is thus primarily aimed at modifying the network topology, 
while thermal fluctuations of the particles about their equilibrium positions 
are ignored (note the contrast with molecular dynamics). Unfortunately, it does 
so at a cost: By deterministically fixing the particle positions at every move 
ergodicity is broken. Hence, the WWW algorithm does {\it not} sample the 
Boltzmann distribution at finite $T$. In its original 
formulation~\cite{citeulike:3269654} this issue was irrelevant, since there the 
goal was simply to generate a well-relaxed (low energy) amorphous network (the 
algorithm thus being merely an optimization tool). However, if one wishes to 
study temperature-driven phase transitions in these systems, then the original 
WWW algorithm needs to be modified. The aim of this paper is to present such a 
modification, and to use it to study the melting transition of a two-dimensional 
3-fold coordinated network.

\section{Model}

To be specific, we consider a two-dimensional $A=L_x \times L_y$ system with 
periodic boundaries containing $i=1,\ldots,N$ particles. To each particle $i$, a 
vector $\vec{r}_i=(x_i,y_i)$ is assigned to denote its position in the plane, as 
well as three integers $\{j,k,l\}$ which denote the labels of the three 
particles to which particle $i$ is bonded. The energy is given by the Keating 
potential~\cite{citeulike:7333784}
\begin{equation}
\label{eq:keating}
\begin{split}
 E = \frac{3\alpha}{16d^2} 
 \sum_{[ij]} \left( \vec{r}_{ij} \cdot \vec{r}_{ij} -d^2 \right)^2 
 \hspace{2cm} \\
 + \, \frac{3\gamma}{8d^2} \sum_{[jik]} \left( \vec{r}_{ij} \cdot 
 \vec{r}_{ik} + \frac{1}{2} d^2 \right)^2,
\end{split}
\end{equation}
with $\vec{r}_{ij} = \vec{r}_j - \vec{r}_i$, and parameters $d=2.35$~\AA, 
$\alpha=2.965$~eV~\AA$^{-2}$, $\gamma=0.285\alpha$ (which are the parameters for 
bulk silicon). The first term is a two-body interaction which equates the 
preferred $i-j$ bond length to $d$; the second term is a three-body interaction 
which sets the preferred $j-i-k$ bond angle to 120 degrees (as appropriate for a 
two-dimensional system). The sums in \eq{eq:keating} extend over bonded 
particles in the connectivity table only. For a system containing $N$ particles, 
there are $3N/2$ two-body terms, and $3N$ three-body terms. The computational 
effort of the energy calculation thus scales linearly with $N$. 

\subsection*{Note on units and conventions}

In what follows, the $NVT$-ensemble is used at density $\rho=N/A=8/(6\sqrt{3} \, 
d^2)$, box aspect ratio $L_x/L_y=3/(2\sqrt{3})$, and $N=8n^2$, with $n$ an 
integer. In this way, the box always \ahum{fits} the groundstate of 
\eq{eq:keating}, i.e.~a perfect honeycomb lattice with lattice constant $d$ (the 
corresponding energy thus being $E=0$). All reported temperatures are given in 
units of ${\rm eV}/k_B$, while free energies (and free energy differences) are 
reported in units of $k_BT$.

\section{The modified bond-switch move}

Consider now a network configuration (i.e.~a set of particle positions and a 
connectivity table) in which a cluster of four particles $c \in \{A,B,C,D\}$ has 
been selected according to the WWW bond-switch move of \fig{fig1}. Let 
$\vec{I}_c$ denote the initial positions of these four particles. Next, perform 
a local energy minimization, whereby only the positions of the four particles in 
the cluster are allowed to change, keeping the positions of all the other 
particles in the network, as well as the connectivity table, fixed. The 
resulting positions are denoted $\vec{P}_c$. The modification of the WWW 
algorithm is based on the observation that, irrespective of the initial 
positions $\vec{I}_c$, the positions $\vec{P}_c$ obtained after local energy 
minimization are always the same. 

This property can be exploited to modify the bond-switch such that both detailed 
balance and ergodicity are obeyed: The cluster $c$ is selected as before; the 
initial positions $\vec{I}_c$, and the positions $\vec{P}_c$ obtained after 
local energy minimization, are recorded. Next, one performs the bond-switch, 
immediately followed by a second local energy minimization where, as before, 
only the particles in the cluster are allowed to move; the resulting positions 
$\vec{Q}_c$ are recorded. Finally, a stochastic process is used to generate four 
random displacements $\vec{\Delta}_c$ around $\vec{Q}_c$, which yield the final 
positions of the four particles: $\vec{Q}_c + \vec{\Delta}_c$. The resulting 
network configuration is accepted with probability
\begin{equation}
\label{eq:vink}
 P_{\rm acc} = \min\left[ 1, 
 \frac{\Pi_c W(\vec{I}_c - \vec{P}_c)}{\Pi_c W(\vec{\Delta}_c)}
 e^{-(E_F - E_I) / k_BT} \right],
\end{equation}
which ensures that detailed balance is maintained. Here, $W(\vec{r})$ is the 
probability that the stochastic process (still to be specified) selects the 
vector $\vec{r}$. Note that, in contrast to \eq{eq:www}, the energies $E_F$ and 
$E_I$ refer to the actual network energy (and not the energy obtained 
after minimization with respect to the particle positions).

Provided the selection process used to generate the displacements 
$\vec{\Delta}_c \equiv (\Delta X_c,\Delta Y_c)$ is ergodic, this algorithm 
faithfully samples the Boltzmann distribution at temperature $T$. Its efficiency 
is set by the details of the selection process. With the particles placed at 
their locally minimized positions $\vec{Q}_c \equiv (X_c,Y_c)$ the energy can be 
approximated as~\cite{note1}
\begin{equation}
\label{eq:taylor}
 E \approx \frac{1}{2} \sum_c 
 \vec{\Delta}_c \cdot {\bf H}_c \cdot \vec{\Delta}_c, \quad
{\bf H}_c = \begin{pmatrix}
E_{X_c X_c} & E_{X_c Y_c} \\
E_{X_c Y_c} & E_{Y_c Y_c} \\
\end{pmatrix},
\end{equation}
where $E_{\alpha\beta}$ is the second derivative of $E$ with respect to $\alpha$ 
and $\beta$ (to be evaluated at the minimized positions). Since \eq{eq:taylor} 
is quadratic in the displacements, the latter are Gaussian distributed in 
thermal equilibrium. A good choice is therefore to draw $\Delta X_c$ from the 
Gaussian probability distribution
\begin{equation}
 W(\Delta X_c) = \frac{1}{\sqrt{2\pi} \sigma_{X,c}} \exp(-\Delta 
 X_c^2/2\sigma_{X,c}^2),
\end{equation}
with $\sigma_{X,c}^2 = k_B T E_{Y_c Y_c} / \det {\bf H}_c$, and similarly for 
$\Delta Y_c$ where $\sigma_{Y,c}^2 = k_B T E_{X_c X_c} / \det {\bf H}_c$. Note 
that, by drawing from a Gaussian distribution, every displacement has a finite 
probability of being selected, and so ergodicity is trivially fulfilled. Since 
the components of $\vec{\Delta}_c$ are selected independently, it holds that 
$W(\vec{\Delta}_c) \propto W(\Delta X_c) \times W(\Delta Y_c)$. The computation 
of the reverse weight $W(\vec{I}_c - \vec{P}_c)$ proceeds analogously, but with 
the derivatives evaluated at $\vec{P}_c$ before the bonds were switched.

The algorithm just described samples the network topology and the particle 
positions simultaneously, which suffices for a simulation. Still, to facilitate 
faster equilibration, it helps to also use a move whereby only the particle 
positions are updated. To this end, one can use the same scheme as above, but 
{\it not} switch the bonds, such that $\vec{P}_c=\vec{Q}_c$ (and so only one 
local energy minimization is required). For this displacement move, it suffices 
to select just a single particle (as opposed to a cluster of four). In the 
results to be presented, both single-particle displacement and WWW moves were 
used, each attempted with equal {\it a priori} probability~\cite{note2}.

\begin{figure}
\begin{center}
\includegraphics[width=0.9\columnwidth]{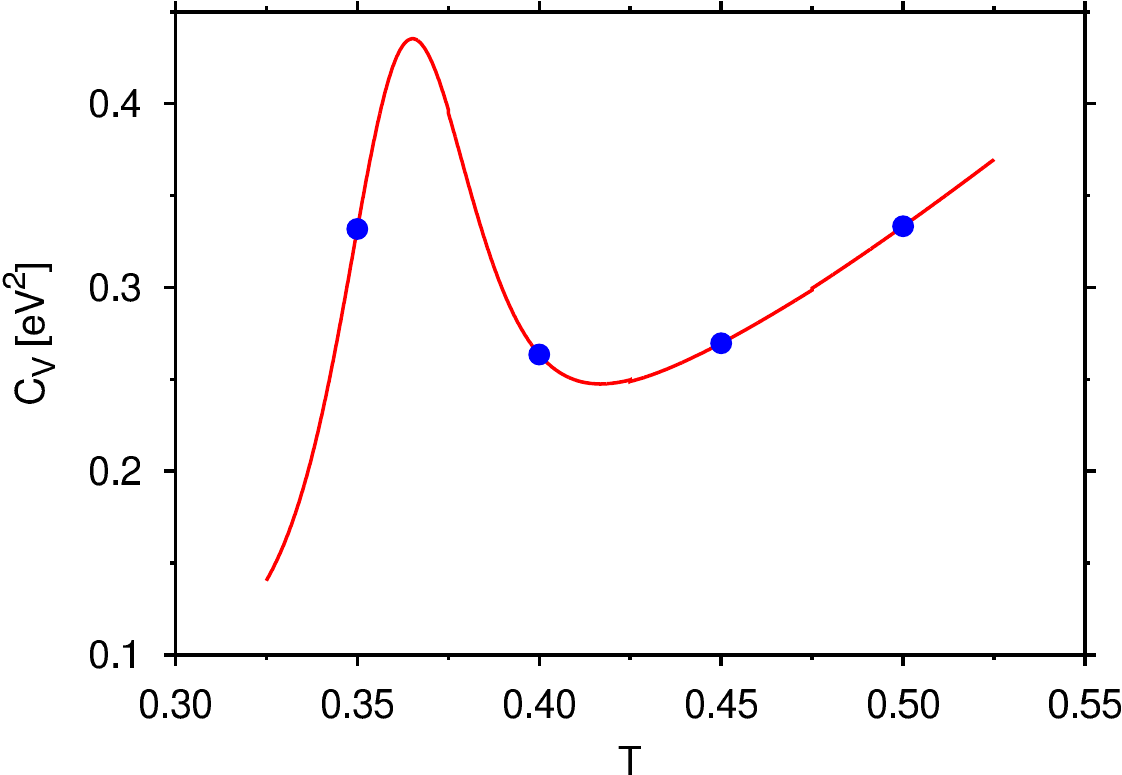}
\caption{\label{fig2} Variation of the specific heat per particle $C_V$ with the 
temperature $T$ as obtained using our algorithm for a network containing $N=128$ 
particles. Simulations were performed at $T = 0.35,0.40,0.45,0.50$ (indicated by 
the dots) with the results of each run subsequently extrapolated over a range 
$\Delta T=\pm 0.05$ using the single histogram 
method~\cite{ferrenberg.swendsen:1988, newman.barkema:1999}.}
\end{center}
\end{figure}

As a first test, we plot in \fig{fig2} how the specific heat per particle 
$C_V=(\avg{E^2}-\avg{E}^2)/N$ of the network varies with the temperature $T$ as 
obtained using our algorithm (with $E$ the Keating energy of \eq{eq:keating} and 
where $\avg{\cdot}$ denotes a thermal average). Simulations were performed at 
four different temperatures (indicated by the dots) and the results of those 
simulations were extrapolated to different temperatures using histogram 
reweighting~\cite{ferrenberg.swendsen:1988, newman.barkema:1999}. The key point 
to note is that the curves of the different runs smoothly \ahum{join-up}, 
providing a strong confirmation that the algorithm is correctly sampling the 
Boltzmann distribution. Incidentally, we observe that the specific heat features 
a maximum: This indicates the melting transition of the network, to be analyzed 
next~\cite{note3}.

\section{Results: Melting of a 2D network}

The melting transition in the Keating model \eq{eq:keating} is from a 
low-temperature ordered phase, to a high-temperature disordered phase. At zero 
temperature, the particles form a perfectly ordered honeycomb lattice, while at 
high temperature the network is spatially disordered (particle positions 
random beyond a certain finite range). Consequently, a phase transition must 
occur, at some finite transition temperature $\tc$. The aim of this section is 
to determine $\tc$, as well as to characterize the transition type (expected to 
be first-order). To this end, we use the hexatic bond-order 
parameter~\cite{citeulike:12820541}
\begin{equation}
\label{eq:bop}
q_6 = \frac{2}{3N} \left| \sum_{[ij]} \exp( \imath 6\theta_{ij}) \right| \quad,
\end{equation}
with the sum over all bonds in the connectivity table (i.e.~a total of $3N/2$ 
terms), and where $\theta_{ij}$ is the angle of the bond between particles $i-j$ 
with respect to an arbitrary reference axes (say, the $x$-axes). In the 
perfectly ordered phase (honeycomb lattice) $q_6=1$, while for the disordered 
phase $\lim_{N \to \infty} q_6=0$ (since \eq{eq:bop} uses the absolute value, 
$q_6$ is never exactly zero, but it approaches this value in a disordered 
system as the system size is increased).

\subsection{Analysis of the free energy}

We will analyze the melting transition via the corresponding order parameter 
distribution $P(q_6)$, defined as the probability to observe a network with 
order parameter $q_6$. The physical significance of $P(q_6)$ is its relation to 
the free energy, $F(q_6) = -k_BT \ln P(q_6)$, providing a convenient means to 
study the phase behavior. To ensure that $P(q_6)$ gets accurately measured, we 
combine our Monte Carlo scheme with a number of high-resolution tools commonly 
used in the study of phase transitions (see Appendix).

\begin{figure}
\begin{center}
\includegraphics[width=0.95\columnwidth]{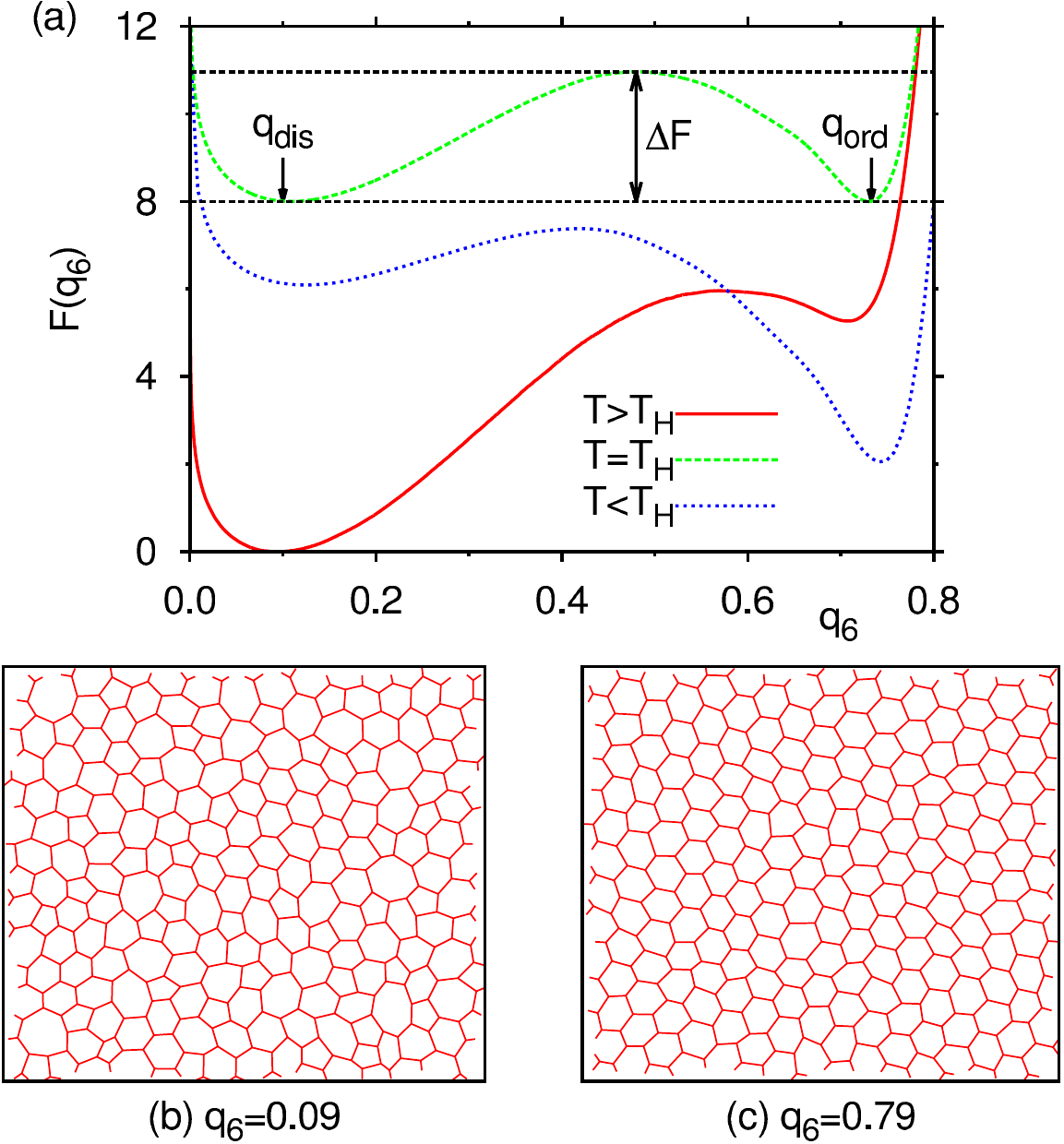}
\caption{\label{free} (a) Typical free energy curves $F(q_6)$ obtained at three 
different temperatures for the Keating model of \eq{eq:keating} using $N=392$ 
particles. The shape variations of $F(q_6)$ are characteristic of a first-order 
phase transition. At $T=T_H$, two-phase coexistence is observed, whereby 
$F(q_6)$ reveals two minima of equal height. The free energy barrier then 
separating the phases is indicated as $\Delta F$. The positions of the minima, 
marked as $q_{\rm dis}$ and $q_{\rm ord}$, reflect the value of the order 
parameter $q_6$ in the disordered and ordered phase, respectively. The snapshots 
show network configurations typical of the disordered (b) and ordered (c) phase 
(vertices represent particle positions; edges represent bonds). All results in 
this figure refer to actual simulation data obtained using the modified 
bond-switch Monte Carlo move of this work.}
\end{center}
\end{figure}

In \fig{free}(a), we show typical free energy curves obtained at three different 
temperatures. At the highest considered temperature, $F(q_6)$ reveals a global 
minimum at a low value of $q_6$, meaning that the disordered phase is the 
thermodynamic stable one (solid curve). A typical snapshot of the disordered 
phase is shown in \fig{free}(b), which reveals a structure containing mostly 
6-fold rings, but with many defects (i.e.~rings that are smaller or larger). At 
low temperature, $F(q_6)$ attains its minimum at a significantly larger value of 
the order parameter (dotted curve). This means that the ordered phase has become 
the stable one. The corresponding snapshot is shown in \fig{free}(c), which 
reveals a rather ordered structure (however, defects do remain; see 
\sect{sec:str}). By tuning the temperature to a special value, $T=T_H$, the two 
minima in the free energy occur at the same height (dashed curve), which marks 
the phase transition. Note that the free energy curves in \fig{free}(a) are 
characteristic of a first-order phase transition, as the global minimum jumps 
discontinuously from $q_{\rm dis}$ to $q_{\rm ord}$ as $T$ is lowered.

\begin{figure*}
\begin{center}
\includegraphics[width=2\columnwidth]{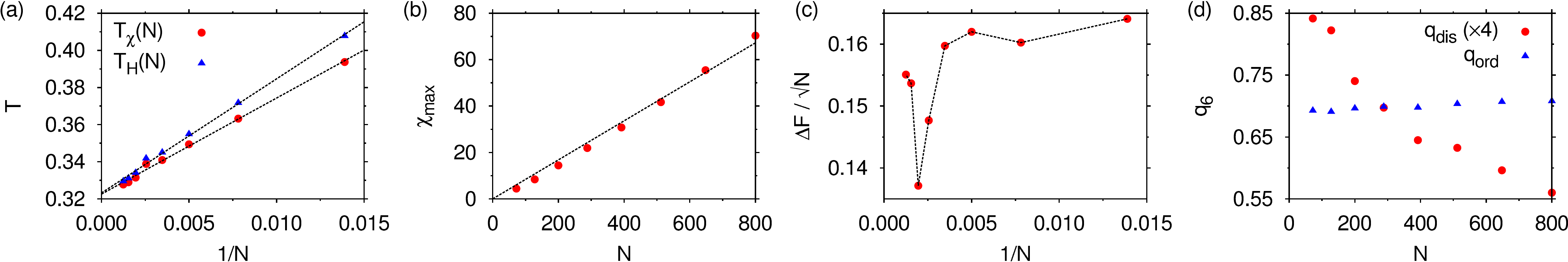}
\caption{\label{fig3} Finite-size scaling analysis of the melting transition in 
the Keating model of \eq{eq:keating} using system sizes $N=72, 128, 200, 288, 
392, 512, 648, 800$. (a) Variation of the pseudo-transition temperatures 
$T_H(N)$ and $T_\chi(N)$ with $1/N$; the lines are linear fits, whose intercepts 
yield $\tc$ of the thermodynamic limit. (b) The susceptibility maximum versus 
$N$. As expected for a first-order transition, a linear increase is revealed. 
(c) The (scaled) free energy barrier $\Delta F/\sqrt{N}$ versus $1/N$; the 
latter is roughly constant (note the fine vertical scale). (d) The locations of 
the free energy minima, $q_{\rm dis}$ and $q_{\rm ord}$, obtained at 
$T=T_\chi(N)$, versus the system size $N$. As expected, $q_{\rm dis}$ of the 
disordered phase decays with $N$, while $q_{\rm ord}$ saturates at a finite 
value. Note that the data for $q_{\rm dis}$ use an enhanced scale.}
\end{center}
\end{figure*}

\subsection{Finite-size scaling analysis}

The temperature $T_H$ where the free energy minima are at equal height depends 
on the size of the system $T_H \equiv T_H(N)$. To accurately locate the 
transition temperature in the thermodynamic limit, $\tc = \lim_{N \to \infty} 
T_H(N)$, requires a finite-size scaling analysis. At a first-order phase 
transition, one expects a shift $\tc - T_H(N) \propto 
1/N$~\cite{citeulike:3716330}. In \fig{fig3}(a), we have plotted $T_H(N)$ versus 
$1/N$ (triangles), which can indeed be fitted quite well with a straight line 
(the intercept of this line yields $\tc$). In addition to $T_H(N)$, it is also 
common to study the finite-size dependence of $T_\chi(N)$, defined as the 
temperature where the susceptibility per particle, $\chi = (\avg{q_6^2} - 
\avg{q_6}^2)/N$, reaches its maximum (thermal averages are trivially computed 
from the normalized order parameter distribution $\avg{q_6^p} = \int_0^1 q_6^p 
P(q_6) dq_6$). The dots in \fig{fig3}(a) show the shift of the latter 
temperature, which also fits quite well to a straight line; combining both 
estimates, we obtain $\tc = 0.323 \pm 0.003$. Also of interest is the value of 
the susceptibility $\chi_{\rm max}$ measured at $T_\chi(N)$. At a first-order 
transition, the latter scales $\propto N$~\cite{citeulike:3716330}, which we 
confirm in \fig{fig3}(b).

Next, we consider the size dependence of the free energy measured at $T=T_H$, 
i.e.~the temperature where the minima are at equal height. As shown in 
\fig{free}(a), the minima are then separated by a free energy barrier $\Delta 
F$, indicated by the vertical double-arrow. At a first order transition, this 
barrier should scale $\propto L^{d-1} \propto \sqrt{N}$, where $L$ denotes the 
linear extension of the system, and $d=2$ the spatial 
dimension~\cite{citeulike:3908342, binder:1982}. The variation of $\Delta 
F/\sqrt{N}$ with $1/N$ is shown in \fig{fig3}(c), which thus should be constant; 
the latter holds to within an uncertainty of $\sim 10$\%. In \fig{fig3}(d), we 
plot the positions of the free energy minima (marked $q_{\rm dis}$ and $q_{\rm 
ord}$ in \fig{free}(a)) as a function of~$N$. Since it is numerically more 
accurate, this analysis was performed at the temperature $T_\chi(N)$ of the 
susceptibility maximum. The positions were obtained using an integral measure
\begin{equation}
 q_{\rm dis} = 2\int_0^c q_6 P(q_6) dq_6, \hspace{2mm}
 q_{\rm ord} = 2\int_c^1 q_6 P(q_6) dq_6, 
\end{equation}
with the \ahum{cut-off} between the phases taken at the average of the full 
distribution $c=\avg{q_6}$. \fig{fig3}(d) shows that $q_{\rm dis}$ decreases 
with $N$, as expected for a disordered phase, while $q_{\rm ord}$ saturates at a 
finite value, consistent with an ordered phase. Note that $q_{\rm ord} \sim 0.7$ 
at the transition, i.e.~distinctly below the value unity of the perfect 
hexagonal lattice. This means that the ordered phase still contains a 
substantial number of defects (see \sect{sec:str}).

All in all, the finite-size scaling results of \fig{fig3} rather strongly 
indicate that the melting transition in the Keating model is first-order, as 
might have been expected.

\begin{figure}
\begin{center}
\includegraphics[width=0.9\columnwidth]{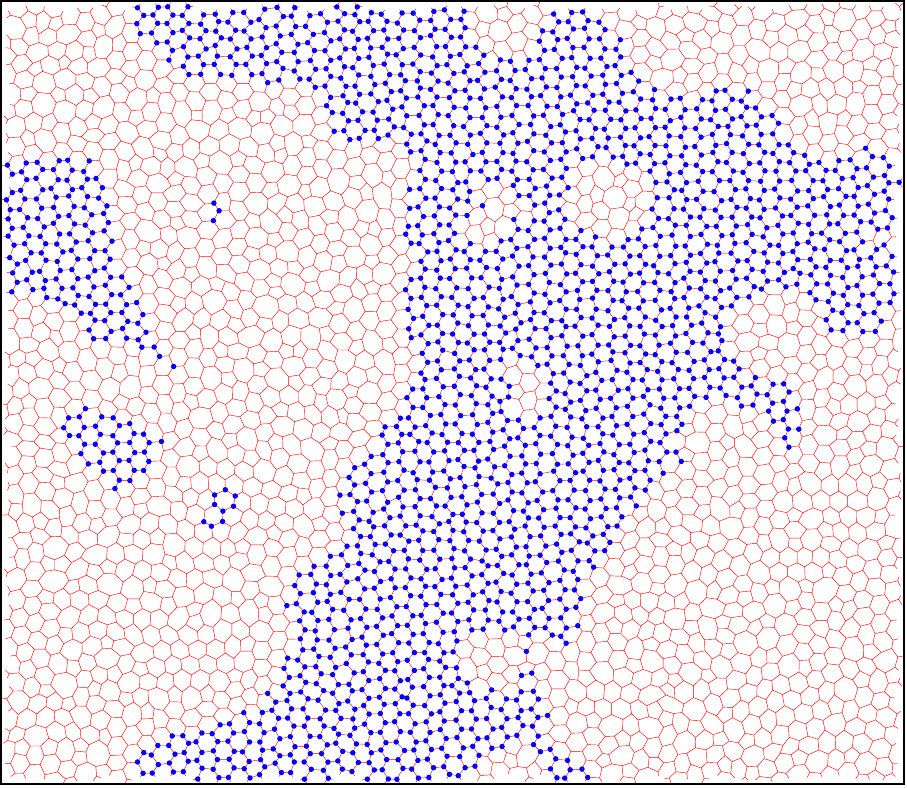}
\caption{\label{5k} Keating network obtained in a simulation where the overall 
order parameter was constrained to a small interval around $q_6=0.4$, which is 
close to the free energy maximum. As temperature, we used $T_\chi(N)=0.3257$ of 
the susceptibility maximum, with $N=5000$. The snapshot strikingly reveals the 
two-phase coexistence that is characteristic of first-order phase transitions. 
As before, edges represent bonds; vertices $i$ whose local order parameter 
$q_{6,i}>0.6$ have been marked with a (blue) dot (for each particle $i$, 
$q_{6,i}$ was computed as in \eq{eq:bop}, but with the sum restricted to 
particles less than five \ahum{steps} away from $i$ in the connectivity table).}
\end{center}
\end{figure}

\subsection{Phase coexistence}

The barrier $\Delta F$ of \fig{free}(a) also has an interesting physical 
interpretation, which we still explore. In order to traverse from one phase to 
the other, a coexistence region must be crossed where both phases appear 
simultaneously. This region will contain a lot of interface, which is the origin 
of the free energy barrier (and thus explains the scaling $\Delta F \propto 
\sqrt{N}$ in two dimensions). The coexistence can be directly visualized if one 
performs a simulation whereby $q_6$ is constrained to a value that is close to 
the maximum of the free energy curve (in practice, one performs such a 
simulation by rejecting those Monte Carlo moves for which $q_6$ strays away too 
much from the desired value). In \fig{5k}, we show a typical snapshot obtained 
via this procedure. As in \fig{free}, the edges represent bonds between 
particles. In addition, those vertices $i$ for which the local order parameter 
$q_{6,i}$ exceeds a certain threshold (and which thus belong to the ordered 
phase) have been marked with a (blue) dot. As the figure strikingly shows, 
virtually all particles that belong to the ordered phase have condensed into one 
large cluster. Since the overall order parameter $q_6$ of the system was chosen 
to be around the maximum of the free energy, this cluster occupies roughly half 
the system area (lever-rule of phase coexistence). In addition, note that the 
cluster has arranged into a \ahum{slab} parallel to one of the edges of the 
simulation cell, as this shape minimizes the length of the interface contour.

\begin{figure}
\begin{center}
\includegraphics[width=0.9\columnwidth]{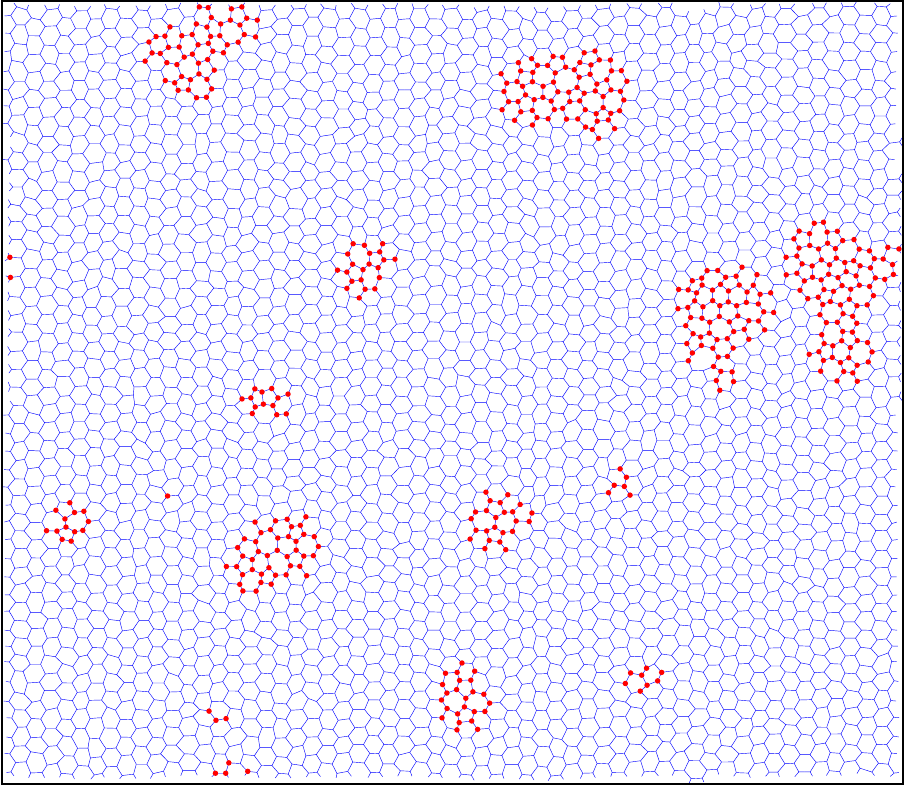}
\caption{\label{5k_low} Keating network obtained in an $NVT$-simulation at 
$T=0.3$, which is well below the transition temperature. An ordered network with 
a finite concentration of defect regions (red dots) is observed. The number of 
particles $N=5000$.}
\end{center}
\end{figure}

\subsection{The structure of the ordered phase}
\label{sec:str}

As announced, we still consider the structure of the ordered phase, i.e.~the 
prevailing phase at low temperature. To this end, we perform an $NVT$-simulation 
at $T=0.3$, which is well below the transition temperature (in this simulation, 
the order parameter was allowed to fluctuate freely). After equilibration, we 
observe that the order parameter saturates at $q_6 \approx 0.735$. In 
\fig{5k_low}, we show a typical snapshot. In this case, the vertices whose local 
order parameter $q_{6,i}<0.6$ have been marked with a (red) dot, which thus 
correspond to disordered regions (edges, as before, represent bonds). We observe 
an overall ordered structure of hexagons, containing \ahum{bubbles} of regions 
that are disordered. In contrast to the snapshot of \fig{5k}, however, these 
bubbles do not coalesce. Hence, the ordered phase is to be regarded as an 
ordered array of hexagons, but with a finite concentration of defect regions. 
The latter reflect thermal fluctuations, that appear at random locations in the 
network. As the temperature is lowered further, these defects are gradually 
frozen out, until, at $T=0$, the perfect honeycomb lattice is reached.

\section{Discussion}

In summary, we have proposed an extension of the bond-switch Monte Carlo move of 
Wooten, Winer, and Weaire~\cite{citeulike:3269654}, in order to correctly 
simulate network forming materials at finite temperature. The algorithm can be 
applied to systems whose potential energy is defined via a connectivity table, 
such as the Keating potential~\cite{citeulike:7333784} (used in this work), or 
the Tu-Tersoff potential for silica~\cite{citeulike:12825492}. A second 
requirement is that, for a given connectivity table, a local energy minimization 
performed on a small cluster of particles always yields the same positions for 
these particles. When these conditions are met, network configurations that 
faithfully sample the Boltzmann distribution are readily generated. This paves 
the way toward the first high-resolution simulations of these materials, since 
all the tools developed for the numerical study of phase transitions 
(finite-size scaling, transition matrix sampling, biased sampling, histogram 
reweighing, and so forth) can now be applied. We have illustrated the merit of 
this approach by performing a detailed analysis of the melting transition in a 
two-dimensional 3-fold coordinated Keating network, which was shown to be 
first-order.

As a future application, an investigation of a silica network, which is a glass 
former, seems particularly fruitful. There are 
indications~\cite{citeulike:12825497} that the dominant mechanism in structural 
relaxation in these materials is, in fact, the bond-switch move of \fig{fig1}. 
In a molecular dynamics simulation, where the natural time scale is 
phonon-based, such bond-switches are rare events. In the present algorithm, the 
natural time scale is event-based, and so it should be easier to probe the 
long-time relaxation regime. We note that \olcite{citeulike:12825497} also 
identified two other relaxation mechanisms, in addition to the bond-switch move 
of \fig{fig1}. Whether these can also be exploited as finite-temperature Monte 
Carlo moves remains to be investigated.

\acknowledgments

We acknowledge financial support by the German research foundation (Emmy Noether 
grant VI~483).

\appendix

\section{Transition matrix sampling}

The aim of transition matrix sampling it to obtain free energy differences based 
on move proposal statistics, rather than on statistics of visited 
states~\cite{citeulike:12476499}. For each attempted Monte Carlo move, two 
matrix elements are updated~\cite{citeulike:2299400}
\begin{equation}
\label{eq:tm}
\begin{split}
M(q_I,q_F) &\leftarrow M(q_I,q_F) + P_{\rm acc}, \\
M(q_I,q_I) &\leftarrow M(q_I,q_I) + 1 - P_{\rm acc},
\end{split}
\end{equation}
with $P_{\rm acc}$ the accept probability of \eq{eq:vink}, $q_I$ the value of 
the order parameter $q_6$ at the start of the move, and $q_F$ that of the 
proposed state (since $q_6$ is continuous, we choose a bin size $\sim 1/N$). The 
update is performed at every attempted move, irrespective of whether it is 
accepted. Hence, even if the accept rate is low, one still collects statistics 
on $M$, which is one advantage of the transition matrix method. In cases where 
the initial and proposed order parameter belong to the same bin, \eq{eq:tm} 
reduces to $M(q_I,q_I) \leftarrow M(q_I,q_I) + 1$.

The matrix elements $M$ are used to estimate the transition probability
\begin{equation}
 T(q_6,q_6') = \frac{ M(q_6,q_6') }{ \sum_x M(q_6,x)} \quad,
\end{equation}
which is the probability that, being in a state with order parameter $q_6$, a 
state with order parameter $q_6'$ is proposed. The latter is related to the free 
energy 
\begin{equation}
 F(q_6) - F(q_6') = 
 k_B T \log \left( \frac{ T(q_6,q_6') }{ T(q_6',q_6) } \right)
 \equiv \Delta(q_6,q_6').
\end{equation}
Hence, during the simulation, a large set of free energy differences 
$\Delta(q_6,q_6')$ is collected. The best estimate of the free energy 
$\tilde{F}(q_6)$ is the one which minimizes the variance
\begin{equation}
\label{eq:var}
 \sum_\Delta w(q_6,q_6') \left( 
 \tilde{F}(q_6) - \tilde{F}(q_6') - \Delta(q_6,q_6') \right)^2 \quad,
\end{equation}
where the sum is over all measured free energy differences. The purpose of 
$w(q_6,q_6')$ is to \ahum{weigh} each measurement $\Delta(q_6,q_6')$ according 
to the magnitude of the corresponding matrix elements $M$ (we choose those 
weights as described in \olcite{citeulike:3577799}). In minimizing \eq{eq:var}, 
one value of $\tilde{F}(q_6)$ is fixed; the remaining values are obtained by 
matrix inversion.

\section{Biased sampling}

To further improve the accuracy of our data, the simulations of the largest 
systems ($N \geq 392$) were performed by adding a bias function $W(q_6)$ (i.e.~a 
function of the order parameter $q_6$) to the Keating energy of \eq{eq:keating}. 
The bias function is chosen such that the simulation visits each value of the 
order parameter with equal probability. That is, in the biased simulations, the 
goal is to observe an order parameter distribution $P(q_6)$ that is flat 
(uniform sampling). In this way, the statistical quality of the data is 
independent of $q_6$ (this property is particularly desirable at first-order 
transitions where otherwise values of $q_6$ corresponding to phase coexistence 
are hardly sampled).

In the biased simulations, the accept probability of the Monte Carlo moves, 
\eq{eq:vink}, is replaced by
\begin{equation*}
 P_{\rm acc}^{\rm bias} = \min \left[1,
 \frac{\Pi_c W(\vec{I}_c - \vec{P}_c)}{\Pi_c W(\vec{\Delta}_c)} 
 e^{-\frac{E_F - E_I}{k_BT} + W(q_I) - W(q_F) } \right],
\end{equation*}
with $q_I \, (q_F)$ the order parameter at the start (end) of the move. To 
obtain uniform sampling, one chooses $W(q_6)=-F(q_6)/k_BT$, where $F(q_6)$ is 
the free energy (which is {\it a priori} unknown). In this work, we first 
performed a non-biased simulation until a sufficiently large range in $q_6$ was 
sampled; the resulting transition matrix data was then used to compute the free 
energy. The biased simulations were subsequently performed using the latter free 
energy as bias function. Note that, in the biased simulations, the transition 
matrix elements are collected exactly as described in Appendix~A, i.e.~using the 
non-biased form of the accept probability, \eq{eq:vink}. In this way, the 
transition matrix elements $M$ of different runs may simply be added (even if 
the runs themselves used different bias functions).

\section{Extrapolations in temperature}

Finally, we explain how the order parameter distribution $P_0(q_6)$, measured at 
temperature $T_0$, is extrapolated to obtain $P_1(q_6)$ at a (nearby) 
temperature $T_1$. To this end, we Taylor expand to second order
\begin{equation}
\label{eq:hrw}
\begin{split}
 \ln P_1(q_6) \approx \ln P_0(q_6) + \Delta\beta
 \left. \frac{d \ln P_0(q_6)}{d\beta} \right|_{\beta=\beta_0} \\ 
 + \frac{1}{2} (\Delta\beta)^2 
 \left. \frac{d^2 \ln P_0(q_6)}{d\beta^2} \right|_{\beta=\beta_0} \quad,
\end{split}
\end{equation}
where $\beta_0=1/k_BT_0$, and $\Delta\beta=1/k_BT_1-\beta_0$. Next, we note that 
$P_0(q_6) = {\rm Tr} \exp(-E/k_BT_0)$, where the trace is over all network 
configurations whose order parameter equals $q_6$, and where $E$ denotes the 
Keating energy. Hence, the first derivative in \eq{eq:hrw} is simply 
$-\avg{E}(q_6)$, i.e.~the negative average value of the Keating energy in the 
bin corresponding to order parameter $q_6$, while the second derivative is the 
energy variance $\avg{E^2}(q_6)-\avg{E}^2(q_6)$ in that bin (both to be measured 
at $T=T_0$). The latter quantities are readily collected in our Monte Carlo 
simulations: At the end of each move, one simply identifies the current order 
parameter bin, and updates the corresponding energy moments. Note that 
\eq{eq:hrw} provides a convenient way to perform extrapolations in 
temperature, without having to store the full energy distribution (only 
the leading two moments are needed).


\begin{thebibliography}{16}
\expandafter\ifx\csname natexlab\endcsname\relax\def\natexlab#1{#1}\fi
\expandafter\ifx\csname bibnamefont\endcsname\relax
  \def\bibnamefont#1{#1}\fi
\expandafter\ifx\csname bibfnamefont\endcsname\relax
  \def\bibfnamefont#1{#1}\fi
\expandafter\ifx\csname citenamefont\endcsname\relax
  \def\citenamefont#1{#1}\fi
\expandafter\ifx\csname url\endcsname\relax
  \def\url#1{\texttt{#1}}\fi
\expandafter\ifx\csname urlprefix\endcsname\relax\def\urlprefix{URL }\fi
\providecommand{\bibinfo}[2]{#2}
\providecommand{\eprint}[2][]{\url{#2}}

\bibitem[{\citenamefont{Wooten et~al.}(1985)\citenamefont{Wooten, Winer, and
  Weaire}}]{citeulike:3269654}
\bibinfo{author}{\bibfnamefont{F.}~\bibnamefont{Wooten}},
  \bibinfo{author}{\bibfnamefont{K.}~\bibnamefont{Winer}}, \bibnamefont{and}
  \bibinfo{author}{\bibfnamefont{D.}~\bibnamefont{Weaire}},
  \bibinfo{journal}{Phys.~Rev. Lett.} \textbf{\bibinfo{volume}{54}},
  \bibinfo{pages}{1392} (\bibinfo{year}{1985}), ISSN \bibinfo{issn}{0031-9007},
  \urlprefix\url{http://dx.doi.org/10.1103/physrevlett.54.1392}.

\bibitem[{\citenamefont{Keating}(1966)}]{citeulike:7333784}
\bibinfo{author}{\bibfnamefont{P.}~\bibnamefont{Keating}},
  \bibinfo{journal}{Phys.~Rev.} \textbf{\bibinfo{volume}{145}},
  \bibinfo{pages}{637} (\bibinfo{year}{1966}), ISSN \bibinfo{issn}{0031-899X},
  \urlprefix\url{http://dx.doi.org/10.1103/physrev.145.637}.

\bibitem[{not({\natexlab{a}})}]{note1}
\bibinfo{note}{An irrelevant additive constant in \eq{eq:taylor} was dropped.
  Note also that \eq{eq:taylor} is not a full second-order Taylor expansion,
  since cross terms ($\propto X_A Y_B, Y_A Y_C$ and so forth) are omitted. This
  does not impede the correctness of the algorithm, but it simplifies the
  numerics, since only $2 \times 2$ matrices need to be dealt with.}

\bibitem[{not({\natexlab{b}})}]{note2}
\bibinfo{note}{The algorithm is completely local (execution time per move
  independent of $N$). Fast convergence of the local energy minimization is
  obtained by updating the particle positions as $\vec{r}_{c,n+1} =
  \vec{r}_{c,n} + \gamma \, {\bf H}_c^{-1} \cdot \vec{f}_{c,n}$, with
  \ahum{fudge factor} $0<\gamma\leq1$, $\vec{r}_{c,n}$ the position of particle
  $c$ at the $n$-th minimization step, and $\vec{f}_{c,n}$ the corresponding
  force acting on the particle (i.e.~the negative gradient of $E$). Due to the
  simplicity of the Keating potential, derivatives can be calculated
  analytically. In cases where the fast minimization scheme did not converge, a
  safer (but slower) steepest descent method was used.}

\bibitem[{\citenamefont{Ferrenberg and
  Swendsen}(1988)}]{ferrenberg.swendsen:1988}
\bibinfo{author}{\bibfnamefont{A.~M.} \bibnamefont{Ferrenberg}}
  \bibnamefont{and} \bibinfo{author}{\bibfnamefont{R.~H.}
  \bibnamefont{Swendsen}}, \bibinfo{journal}{Phys.~Rev. Lett.}
  \textbf{\bibinfo{volume}{61}}, \bibinfo{pages}{2635} (\bibinfo{year}{1988}),
  \urlprefix\url{http://dx.doi.org/10.1103/physrevlett.61.2635}.

\bibitem[{\citenamefont{Newman and Barkema}(1999)}]{newman.barkema:1999}
\bibinfo{author}{\bibfnamefont{M.~E.~J.} \bibnamefont{Newman}}
  \bibnamefont{and} \bibinfo{author}{\bibfnamefont{G.~T.}
  \bibnamefont{Barkema}}, \emph{\bibinfo{title}{Monte Carlo Methods in
  Statistical Physics}} (\bibinfo{publisher}{Clarendon Press},
  \bibinfo{address}{Oxford}, \bibinfo{year}{1999}),
  \urlprefix\url{http://books.google.de/books?id=J5aLdDN4uFwC\&hl=en}.

\bibitem[{not({\natexlab{c}})}]{note3}
\bibinfo{note}{Around $\tc$, the accept rate of the (single particle)
  displacement moves $\sim 95$\%, while that of the bond-switch moves $\sim
  0.02$\% (i.e.~significantly lower). Since the algorithm is local the
  computational effort per move is small. Our implementation (currently a
  single processor version) running on an Intel Xeon E5-2660 at 2.2~GHz reaches
  $\sim 1.3$ million attempted moves per minute.}

\bibitem[{\citenamefont{Nelson and Halperin}(1979)}]{citeulike:12820541}
\bibinfo{author}{\bibfnamefont{D.}~\bibnamefont{Nelson}} \bibnamefont{and}
  \bibinfo{author}{\bibfnamefont{B.}~\bibnamefont{Halperin}},
  \bibinfo{journal}{Phys.~Rev. B} \textbf{\bibinfo{volume}{19}},
  \bibinfo{pages}{2457} (\bibinfo{year}{1979}), ISSN \bibinfo{issn}{0163-1829},
  \urlprefix\url{http://dx.doi.org/10.1103/physrevb.19.2457}.

\bibitem[{\citenamefont{Challa et~al.}(1986)\citenamefont{Challa, Landau, and
  Binder}}]{citeulike:3716330}
\bibinfo{author}{\bibfnamefont{M.~S.~S.} \bibnamefont{Challa}},
  \bibinfo{author}{\bibfnamefont{D.~P.} \bibnamefont{Landau}},
  \bibnamefont{and} \bibinfo{author}{\bibfnamefont{K.}~\bibnamefont{Binder}},
  \bibinfo{journal}{Phys.~Rev. B} \textbf{\bibinfo{volume}{34}},
  \bibinfo{pages}{1841} (\bibinfo{year}{1986}),
  \urlprefix\url{http://dx.doi.org/10.1103/physrevb.34.1841}.

\bibitem[{\citenamefont{Lee and Kosterlitz}(1991)}]{citeulike:3908342}
\bibinfo{author}{\bibfnamefont{J.}~\bibnamefont{Lee}} \bibnamefont{and}
  \bibinfo{author}{\bibfnamefont{J.~M.} \bibnamefont{Kosterlitz}},
  \bibinfo{journal}{Phys.~Rev. B} \textbf{\bibinfo{volume}{43}},
  \bibinfo{pages}{3265} (\bibinfo{year}{1991}),
  \urlprefix\url{http://dx.doi.org/10.1103/physrevb.43.3265}.

\bibitem[{\citenamefont{Binder}(1982)}]{binder:1982}
\bibinfo{author}{\bibfnamefont{K.}~\bibnamefont{Binder}},
  \bibinfo{journal}{Phys. Rev. A} \textbf{\bibinfo{volume}{25}},
  \bibinfo{pages}{1699} (\bibinfo{year}{1982}),
  \urlprefix\url{http://dx.doi.org/10.1103/PhysRevA.25.1699}.

\bibitem[{\citenamefont{Tu and Tersoff}(2000)}]{citeulike:12825492}
\bibinfo{author}{\bibfnamefont{Y.}~\bibnamefont{Tu}} \bibnamefont{and}
  \bibinfo{author}{\bibfnamefont{J.}~\bibnamefont{Tersoff}},
  \bibinfo{journal}{Phys.~Rev. Lett.} \textbf{\bibinfo{volume}{84}},
  \bibinfo{pages}{4393} (\bibinfo{year}{2000}), ISSN \bibinfo{issn}{0031-9007},
  \urlprefix\url{http://dx.doi.org/10.1103/physrevlett.84.4393}.

\bibitem[{\citenamefont{Barkema and Mousseau}(1998)}]{citeulike:12825497}
\bibinfo{author}{\bibfnamefont{G.}~\bibnamefont{Barkema}} \bibnamefont{and}
  \bibinfo{author}{\bibfnamefont{N.}~\bibnamefont{Mousseau}},
  \bibinfo{journal}{Phys.~Rev. Lett.} \textbf{\bibinfo{volume}{81}},
  \bibinfo{pages}{1865} (\bibinfo{year}{1998}), ISSN \bibinfo{issn}{0031-9007},
  \urlprefix\url{http://dx.doi.org/10.1103/physrevlett.81.1865}.

\bibitem[{\citenamefont{Fitzgerald et~al.}(1999)\citenamefont{Fitzgerald,
  Picard, and Silver}}]{citeulike:12476499}
\bibinfo{author}{\bibfnamefont{M.}~\bibnamefont{Fitzgerald}},
  \bibinfo{author}{\bibfnamefont{R.~R.} \bibnamefont{Picard}},
  \bibnamefont{and} \bibinfo{author}{\bibfnamefont{R.~N.}
  \bibnamefont{Silver}}, \bibinfo{journal}{EPL} p. \bibinfo{pages}{282}
  (\bibinfo{year}{1999}),
  \urlprefix\url{http://dx.doi.org/10.1209/epl/i1999-00257-1}.

\bibitem[{\citenamefont{Errington}(2003)}]{citeulike:2299400}
\bibinfo{author}{\bibfnamefont{J.~R.} \bibnamefont{Errington}},
  \bibinfo{journal}{J.~Chem.~Phys.} \textbf{\bibinfo{volume}{118}},
  \bibinfo{pages}{9915} (\bibinfo{year}{2003}),
  \urlprefix\url{http://dx.doi.org/10.1063/1.1572463}.

\bibitem[{\citenamefont{Shell et~al.}(2003)\citenamefont{Shell, Debenedetti,
  and Panagiotopoulos}}]{citeulike:3577799}
\bibinfo{author}{\bibfnamefont{M.~S.} \bibnamefont{Shell}},
  \bibinfo{author}{\bibfnamefont{P.~G.} \bibnamefont{Debenedetti}},
  \bibnamefont{and} \bibinfo{author}{\bibfnamefont{A.~Z.}
  \bibnamefont{Panagiotopoulos}}, \bibinfo{journal}{J.~Chem.~Phys.}
  \textbf{\bibinfo{volume}{119}}, \bibinfo{pages}{9406} (\bibinfo{year}{2003}),
  \urlprefix\url{http://dx.doi.org/10.1063/1.1615966}.

\end{thebibliography}
\end{document}